\documentclass{jfm}
\usepackage{tabularx}
\usepackage{graphicx}
\usepackage{natbib}
\usepackage{pifont}
\usepackage{soul}
\usepackage{amsmath}
\usepackage{amsfonts}
\usepackage{color}
\usepackage{mathtools}
\usepackage{relsize}
\usepackage{wasysym}
\usepackage[hidelinks=true]{hyperref}

\include{Header}

\newcommand{\vol}{\mathop{\ooalign{\hfil$V$\hfil\cr\kern0.08em--\hfil\cr}}\nolimits}

\newcommand{\bnabla}{\boldsymbol{\nabla}}

\newcommand{\mbfi}[1]{\mathsfbi{#1}}
\newcommand{\bs}[1]{\boldsymbol{#1}}

\newcommand{\norm}[1]{{\left\lVert#1\right\rVert}_2}
\newcommand{\tm}[1]{\left\langle#1\right\rangle}
\newcommand{\mrm}[1]{\mathrm{#1}}

\title[Data-driven modelling of turbulent convection using DMD-enhanced FDT]{Data-driven reduced modelling of turbulent Rayleigh-B\'enard convection using DMD-enhanced Fluctuation-Dissipation Theorem}

\author[M. A. Khodkar and P. Hassanzadeh]{M. A. Khodkar$^1$ and Pedram Hassanzadeh$^{1,2}$\thanks{Email addresses for correspondence: mkhodkar@rice.edu and pedram@rice.edu}}

\affiliation{$^1$Department of Mechanical Engineering, Rice University, Houston, TX\\
$^2$Department of Earth, Environmental, and Planetary Sciences, Rice University, Houston, TX}

\date{\today}

\pubyear{2018}
\volume{??}
\pagerange{???--???}
\date{?; revised ?; accepted ?. - To be entered by editorial office}

\begin{document}

\maketitle

\begin{abstract}
A data-driven, model-free framework is introduced for calculating Reduced-Order Models (ROMs) capable of accurately predicting time-mean responses to external forcings, or forcings needed for specified responses, e.g., for control, in fully turbulent flows. The framework is based on using the Fluctuation-Dissipation Theorem (FDT) in the space of a limited number of modes obtained from Dynamic Mode Decomposition (DMD). Using the DMD modes as the basis functions, rather than the commonly used Proper Orthogonal Decomposition (POD) modes, resolves a previously identified problem in applying FDT to high-dimensional, non-normal turbulent flows. Employing this DMD-enhanced FDT method (FDT$_\mrm{DMD}$), a linear ROM with horizontally averaged temperature as state vector, is calculated for a 3D Rayleigh-B\'enard convection system at the Rayleigh number of $10^6$ using data obtained from Direct Numerical Simulation (DNS). The calculated ROM performs well in various tests for this turbulent flow, suggesting FDT$_\mrm{DMD}$ as a promising method for developing ROMs for high-dimensional, turbulent systems.


\end{abstract}



\section{Introduction \label{section:Intro}}

Developing accurate Reduced-Order Models (ROMs) for high-dimensional and complex turbulent systems is the subject of ever-growing interest and extensive research \citep{mezic2013analysis,rowley2017model}. For example, reduced-order modelling of buoyancy-driven turbulence, which is prevalent in many engineering flows (e.g., energy systems) and natural flows (e.g., atmospheric/ocean circulations), has been actively pursued by the fluid dynamics and climate science communities in the past few decades; see below, also \citet{Khodkar2018} and \citet{Tu2014} and references therein.  

In many reduced-order modelling efforts, an alternative to the computationally prohibitive high-dimensional systems of the nonlinear partial differential equations governing the turbulent fluid flow is sought in the form of low-dimensional systems of Ordinary Differential Equations (ODEs), such as the linear ROM    
\begin{eqnarray}
	 \dot{\bs{x}}(t) = \mbfi{L} \, \bs{x}(t) + \bs{f}(t) \, . \label{eqn:Intro1}
\end{eqnarray}
Here $\bs{x}$ and $\mathsfbi{L}$ are, respectively, the system's state vector and the evolution operator or linear response function. $\bs{f}(t)$ may include external forcings/actuations (e.g., controlling inputs) and stochastic representation of unresolved scales/physics. Calculating accurate $\mbfi{L}$ for high-dimensional, nonlinear systems such as fully turbulent flows using {\it data-driven} methods is the goal of many reduced modelling studies, including the present one.

In recent years, significant efforts, particularly in the fluid dynamics community, have been focused on calculating $\mbfi{L}$ using some variant of Dynamic Mode Decomposition (DMD) \citep[e.g.,][]{Schmid2010,Rowley2009,Tu2014,williams2015data,Brunton2017chaos,Arbabi2017,korda2018convergence}, which provides a {\it finite-dimensional}, data-driven approximation (see \S\ref{section:Theory}) to the system's Koopman operator, which is {\it infinite-dimensional} \citep{koopman1931hamiltonian,mezic2005spectral}. DMD-based methods have been applied to a variety of fluid flows \citep{mezic2013analysis,Tu2014,rowley2017model}, including buoyancy-driven turbulence \citep[e.g.,][]{Kramer2017}. Although these studies have produced promising results, particularly not far from the onset of linear instability, application of these methods to fully turbulent flows is currently the subject of extensive research.

In climate science, the focus has been mainly on using the Fluctuation-Dissipation Theorem (FDT) \citep{Leith1975,majda2005information}. FDT, a powerful tool from statistical physics \citep{nyquist1928thermal,kubo1966fluctuation}, provides a data-driven approximation of $\mbfi{L}$ for nonlinear systems from the Fokker-Planck equation, see \S\ref{section:Theory}. The $\mbfi{L}$  calculated using FDT ($\mbfi{L}_\mathrm{FDT}$ hereafter) is of particular interest because it is, theoretically, expected to predict long-time-mean responses to external forcings or forcings needed for a specified mean response in nonlinear systems via Eqs.~(\ref{eqn:Intro1}) \citep{majda2005information}. FDT has been found to work well when applied to very simple models of geophysical turbulence such as the Lorenz equations, however, calculating accurate $\mbfi{L}$ for more complex systems such as the quasi-geostrophic equations or large-scale atmospheric turbulence has been found challenging \citep{gritsun2007climate,Cooper2011,fuchs2015exploration,lutsko2015applying,Hassanzadeh2016b}. The latter study showed that a commonly used step that involves employing the leading (orthogonal) modes obtained from Proper Orthogonal Decomposition (POD) as basis functions for truncating the data can lead to significant inaccuracy in $\mbfi{L}$ if the system is non-normal, which is common in geophysical flows. This step is necessary when the dataset is {\it short} as is often the case for high-dimensional systems, {see \S\ref{section:Theory} for further discussions}.   

\begin{figure}
  \centerline{\includegraphics[width=0.51\textwidth]{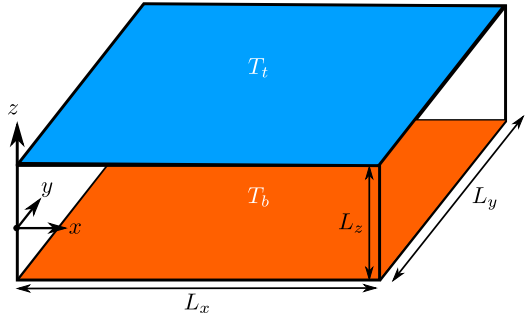}}
  \caption{The 3D Rayleigh-B\'{e}nard Convection (RBC) system. Temperature at top (bottom) wall is held at $T_t$ ($T_b$) and $\Delta T = T_b - T_t > 0$. The horizontal directions ($x$--$y$) are periodic. The no-slip boundary condition is enforced at the walls. $L_x=L_y=\pi L_z$, $Pr = 0.707$, and $Ra = 10^6$.}
\label{fig:Config}
\end{figure}

As a result, it is worthwhile to further examine the performance of $\mbfi{L}_\mathrm{FDT}$  in the context of a canonical, fully turbulent flow system and explore whether basis functions other than POD modes can improve the performance of FDT for developing ROMs for high-dimensional systems. Along these lines, the purpose of this study is twofold: 

\begin{enumerate}
\item[1)] To examine the performance of FDT in calculating $\mbfi{L}$ for a fully turbulent flow, i.e., the 3D Rayleigh-B\'enard Convection (RBC) at the Rayleigh number of $Ra=10^6$ (figure~\ref{fig:Config}). Direct Numerical Simulation (DNS) of RBC, a fitting prototype for buoyancy-driven turbulence, is used to generate the data for FDT. 
\item[2)] To show that using DMD modes, rather than the commonly used POD modes (also known as EOF modes), as the basis functions in the FDT calculation can resolve the problem  previously identified in \citet{Hassanzadeh2016b} and remarkably improve the performance of FDT applied to high-dimensional, turbulent systems.    
\end{enumerate}


Furthermore, this work aims to better connect the seemingly independent advances in the fluid dynamics and climate science communities. It is worth mentioning here, and further discussing in \S\ref{section:Theory}, that FDT and DMD are not unrelated. In fact, another method, called Linear Inverse Modelling (LIM, \citet{penland1989random}) that is also derived from the Fokker-Planck equation and is closely related to FDT, is, as pointed out in \citet{Tu2014}, mathematically equivalent to DMD, although LIM and DMD are derived using different concepts. These connections are not surprising given that the Koopman operator is the adjoint of the Perron-Ferbenius operator \citep{klus2016numerical}, and that the latter is connected to the Fokker-Planck equation \citep{lasota2013chaos}.

This paper is structured  as follows. The formulations of FDT and DMD are discussed in \S\ref{section:Theory}. The 3D RBC system, its ROM, and the DNS solver are described in  \S\ref{section:DNS}. In \S\ref{section:Results}, the accuracy of $\mbfi{L}$ in predicting the time-mean response to forcing is examined for FDT with basis functions of POD modes (FDT$_\mrm{POD}$) and DMD modes (FDT$_\mrm{DMD}$) using DNS of RBC and Stochastic ODEs (SDEs). Summary and future work are discussed in \S\ref{section:Conclusion}.

\color{black}


\section{Fluctuation-Dissipation Theorem (FDT) and DMD \label{section:Theory}}
Let $\bs{x}_t \in \mathbb{R}^m$ be time-mean-removed measurements (e.g, from DNS data) of the state vector (which might involve velocity, temperature) over $m$ grid points at time $t$, and
\begin{eqnarray}
\mbfi{X}_o = \{\bs{x}_{\Delta t} \; \; \bs{x}_{2\Delta t} \; \ldots \; \bs{x}_{N\Delta t} \} \; \; \; \; , \; \; \; \; \mbfi{X}_\tau = \{\bs{x}_{\tau+\Delta t} \;  \; \bs{x}_{\tau+2\Delta t} \; \ldots \; \bs{x}_{\tau+N\Delta t} \} \, ,
\label{eqn:X}
\end{eqnarray}
where $\Delta t$ is the sampling interval and $N$ is the number of samples. {Therefore, $\mbfi{X_0}$ and $\mbfi{X_{\tau}}$ are $m \times N$ matrices.} Below we present the mathematical formulation and numerical procedure for calculating $\mbfi{L}$ from matrices like $\mbfi{X}_o$ and $\mbfi{X}_\tau$ using FDT, LIM, and DMD. It is more convenient to start with the latter.

\subsection{Dynamic Mode Decomposition (DMD) and Linear Inverse Modelling (LIM)}\label{section:DMD}
Following the Exact DMD formulation of \citet{Tu2014}, operator $\mbfi{A}_\mrm{DMD} = \exp\left(\mbfi{L}_\mrm{DMD} {\tau} \right)$ is calculated as
\begin{eqnarray}
\mbfi{A}_\mrm{DMD} = \mbfi{X}_{\tau} \mbfi{X}_o^+ \, .
\label{eqn:DMD}
\end{eqnarray}
Here $+$ denotes the pseudoinverse. The DMD modes (values) are the eigenvectors (values) of $\mbfi{A}_\mrm{DMD} $. In practice, one often uses $\tau=\Delta t$ and calculates the reduced Singular Value Decomposition (SVD) $\mbfi{X}_o= \mbfi{U} \, \mbfi{S} \, \mbfi{V}^\dag $ and then $\mbfi{A}_\mrm{DMD} = \mbfi{U}^\dag \mbfi{X}_\tau \mbfi{V} \, \mbfi{S}^{-1}$ where $\dag$ denotes the adjoint. {Note that some of the DMD modes might be complex; when we later choose a subset of DMD modes as basis functions, we ensure that the complex conjugate of any chosen complex DMD mode is also included (see \S\ref{sec:FDTDMD} for more details).}

\citet{penland1989random} showed that operator $\mbfi{A}_\mrm{LIM} = \exp\left(\mbfi{L}_\mrm{LIM} {\tau} \right)$ can be calculated, from the Fokker-Planck equation, as
\begin{eqnarray}
\mbfi{A}_\mrm{LIM} = \mbfi{C}_{\tau} \mbfi{C}_o^{-1} \, ,
\label{eqn:LIM}
\end{eqnarray}
where $\mbfi{C}_{\tau}=\mbfi{X}_{\tau} \mbfi{X}_o^\dag$ is lag-$\tau$ covariance matrix. {In practice, for high-dimensional systems, covariance matrices are nearly singular (i.e., ill-conditioned), because often there is not enough data available to accurately characterize the system that has a large number of degrees of freedom (i.e., the dataset is short) . To overcome this problem, a common regularization strategy is to first project $\bs{x}_t$ onto the leading $r$ POD/EOF modes (obtained from SVD of $\mbfi{X}_o$) and perform the calculations in Eqn.~(\ref{eqn:LIM}) in this reduced space}. $r$ is chosen such that the retained POD modes represent at least $95\%$ (or even $99\%$) of the variance \citep{penland1989random,Ring2008,lutsko2015applying}. 


Note that because $\mbfi{X}_{\tau} \mbfi{X}_o^+=\mbfi{X}_{\tau} (\mbfi{X}_o^\dag (\mbfi{X}_o \mbfi{X}_o^\dag)^{-1})=\mbfi{C}_{\tau} \mbfi{C}_o^{-1}$, Eqns.~(\ref{eqn:DMD}) and (\ref{eqn:LIM}) are equivalent; see \citet{Tu2014} for further discussions. It should be pointed out that the Koopman operator, which describes the evolution of {\it observables}, and Perron-Frobenius operator, which describes the {\it transition density function}, are adjoints \citep[e.g.,][]{klus2016numerical}, and that if the stochastic noise vanishes, the Fokker-Planck operator reduces to the Perron-Frobenius operator; see e.g., \citet[][chp.~11]{lasota2013chaos} and \citet{giannakis2017data}.

\subsection{Fluctuation-Dissipation Theorem (FDT)}
According to FDT \citep{kubo1966fluctuation,Leith1975}, the linear response function can be calculated as
\begin{eqnarray}
	 \mbfi{L}_\mrm{FDT} = -\bigg[\int_0^{\infty} \mbfi{C}_{\tau} \mbfi{C}_0^{-1} \, \mrm{d} \tau \bigg]^{-1} \, . \label{eqn:FDT}
\end{eqnarray}
Note that the integrand is basically $\mbfi{A}_\mrm{DMD}$ or $\mbfi{A}_\mrm{LIM}$, consistent with integrating $\mbfi{A} = \exp\left(\mbfi{L} {\tau} \right)$ over $\tau$ from $0$ to $\infty$ if $\mbfi{L}$ only has decaying modes. {For details of the derivation of Eqn.~(\ref{eqn:FDT}) from the Fokker-Planck equation see} \citet{majda2005information} and \citet{gritsun2007climate}. Finding $\mbfi{L}_\mrm{FDT}$ is of particular interest, because it allows calculating the time-mean response to an imposed forcing or the forcing needed for a specified response via  $ \mbfi{L}_\mrm{FDT}  \langle \bs{x} \rangle = - \langle \bs{f} \rangle$ where $\langle \, \rangle$ denotes long-time averaging.  It should be noted that the key underlying assumptio {for the applicability of $\mbfi{L}_\mrm{FDT}$} is not that the system is linear, but that the forcing is weak enough such that the response of the nonlinear system changes linearly with the forcing \citep{gritsun2007climate,Cooper2011}.

In practice, similar to LIM, the calculations required to evaluate the expression appearing in Eqn.~(\ref{eqn:FDT}) are typically done in the reduced space of the leading $r$ POD modes to avoid singular covariance matrices. The upper bound of the integral is also replaced with a finite limit $\tau_\infty$.  The reason(s) behind the inaccuracy of $\mbfi{L}_\mrm{FDT} $ calculated for high-dimensional systems (see \S\ref{section:Intro}) is not fully understood, and often attributed to a number of potential fundamental and practical issues. For example, Eqn.~(\ref{eqn:FDT}) (and \ref{eqn:LIM}) is exact only if the statistics of $\bs{x}$ is Gaussian \citep{majda2005information,gritsun2007climate}, which is not the case for turbulent flows such as atmospheric circulation \citep{Cooper2011,hassanzadeh2014responses}. Examples of practical issues include unsuitable choice of $r$ or $\tau_\infty$, short dataset, and shortcomings of POD modes as basis functions \citep{Cooper2013,Hassanzadeh2016b}. The latter issue is particularly significant and is addressed in the current study. But first, we describe in \S\ref{section:DNS} the DNS dataset that is used for calculating matrices in (\ref{eqn:X}).



\color{black}

\section{The 3D RBC Mathematical Model, DNS Solver \& 1D ROM \label{section:DNS}}




The RBC system of figure~\ref{fig:Config} is modeled using the 3D Boussinesq equations. Choosing the height $L_z$, temperature $\Delta T=T_b - T_t$, and diffusive time scale $\tau_{diff}=L^2_z/\kappa$ ($\kappa$ is the thermal diffusivity) as characteristic scales, the dimensionless equations are
\begin{eqnarray}
\bnabla^* \cdot \boldsymbol{u}^* &=& 0 \label{eq:div} \, , \\
\frac{\partial \boldsymbol{u}^*}{\partial t^*}+(\boldsymbol{u}^* \cdot \bnabla^*)  \boldsymbol{u}^*  &=& - \bnabla^* p^* + {Pr} \, \nabla^{*2} \boldsymbol{u}^* + {Ra \, Pr} \,(T^*-T^*_{cond}) \hat{\mathbf{e}}_z  \label{eq:mom} \, , \\
\frac{\partial {T}^*}{\partial t^*}+(\boldsymbol{u}^* \cdot \bnabla^*) {T}^*  &=&  \nabla^{*2} {T}^*  \, , \label{eq:temp}
\end{eqnarray}
where $\boldsymbol{u}^*$, $T^*$, and $T^*_{cond}=1/2-z^*$ are the 3D velocity field, temperature, and conduction temperature profile, respectively. The superscript $*$ indicates dimensionless variables and operators. We define Rayleigh and Prandtl numbers as $Ra = (g \alpha \Delta T L^3_z)/(\nu \kappa)$ and $Pr = {\nu}/{\kappa}$, where $g$, $\alpha$, and $\nu$ are the gravitational acceleration and fluid's thermal expansion coefficient and kinematic viscosity, respectively. 

Equations (\ref{eq:div})--(\ref{eq:temp}) for the system shown in figure~\ref{fig:Config} are simulated (at $Ra = 10^6$) using a pseudo-spectral Fourier-Fourier-Chebyshev DNS solver with the resolution of $128 \times 128 \times 129$ (see \citet{Khodkar2018} for more details). The spatio-temporal analysis of the DNS data in \citet{Khodkar2018} shows that the flow is fully turbulent at this $Ra$, which is around $585$ times higher than the critical $Ra$ for the onset of linear instability. 


As shown in \citet{Khodkar2018}, a 1D ROM in the form of Eqn.~(\ref{eqn:Intro1}) can be formulated for this 3D RBC system:
\begin{equation}
	 \dot{\overline{\theta}} = \mbfi{L} \, \overline{\theta} + \overline{f}    \, , \label{eqn:Theory7}
\end{equation}
where the overbar indicates horizontal ($x$--$y$) averaging. Here the state vector $\bs{x}=\overline{\theta}(z,t)$ is the response of horizontally averaged temperature to external forcing $\overline{f}(z,t)$, i.e., deviation from long-time-mean, horizontally averaged temperature of the unforced system (hereafter, ``unforced'' systems refer to (\ref{eq:div})--(\ref{eq:temp}); in the forced systems, an {\it external} forcing $f$ is added to (\ref{eq:temp})). {Here ``1D'' highlights that the ROM describes the system in only one spatial direction, $z$, as the state vector is horizontally averaged.} $\mbfi{L}$ in (\ref{eqn:Theory7}) includes the vertical heat flux by molecular diffusion as well as vertical {\it eddy} heat flux. Because of the latter, $\mbfi{L}$ cannot be derived directly from (\ref{eq:div})--(\ref{eq:temp}). \citet{Khodkar2018} showed that $\mbfi{L}$ can be accurately calculated using the Green's function (GRF) method \citep{kuang2010linear,Hassanzadeh2016a}, which requires many forced DNS of Eqns.~(\ref{eq:div})--(\ref{eq:temp}). As demonstrated in \S\ref{section:Results}, using FDT, $\mbfi{L}$ in (\ref{eqn:Theory7}) can be accurately calculated  from a dataset obtained from unforced DNS.


\color{black}

\section{Results \label{section:Results}}
\subsection{DNS of RBC at $Ra=10^6$: FDT$_\mrm{POD}$}

\begin{figure}
  \centerline{\includegraphics[width=0.95\textwidth]{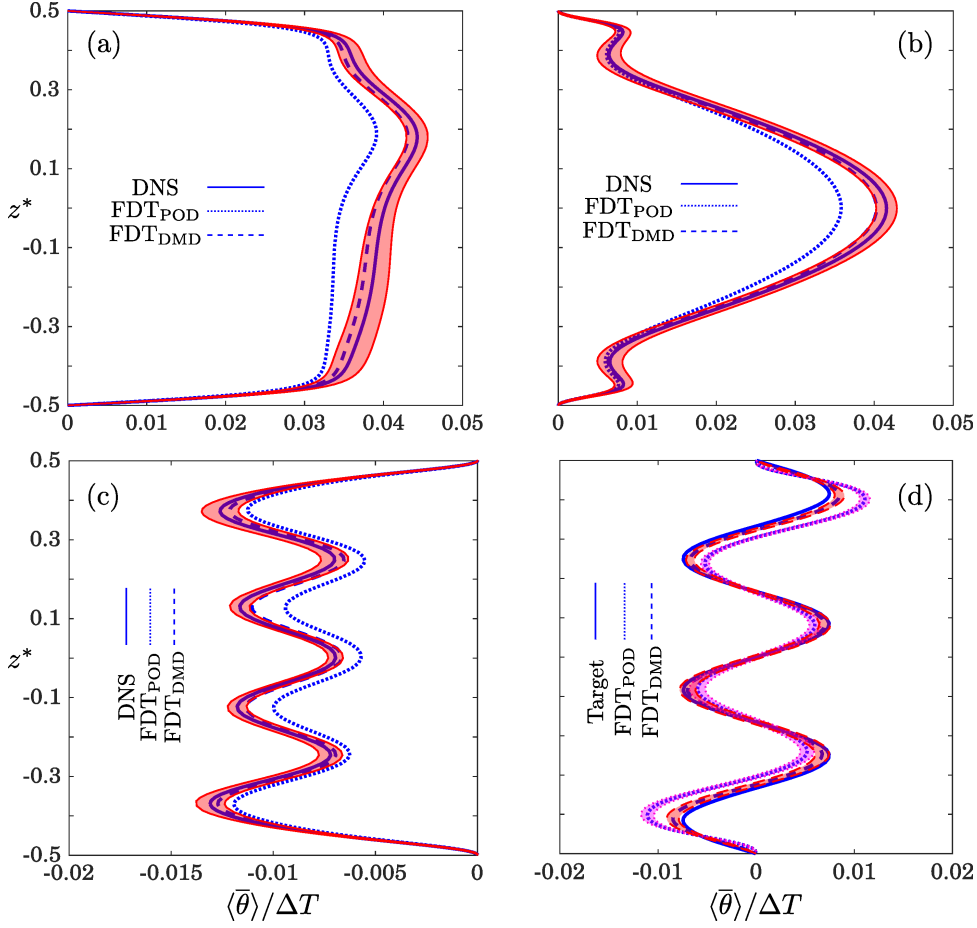}}
  \caption{Time-mean responses $\langle \overline{\theta} \rangle$ to forcings in the form of $f_i = (\Delta T / \tau_{diff}) \hat{f}_i$: (a) $\hat{f_1} = 10 \exp[-(z^* - 0.2)^2/0.1^2]$, (b) $\hat{f_2} = 20 \cos(2\pi z^*)$, (c) $\hat{f_3} = 20 \cos(8\pi z^*)$. To quantify the ``true'' responses and their uncertainty, each long, forced DNS dataset is divided into 8 equal segments and solid lines (red shadings) show their mean ($\pm 1$ standard deviation).  Dotted (dashed) lines show $-\mbfi{L}_\mrm{FDT_{\mrm{POD}}}^{-1} f_i$ ($-\mbfi{L}_\mrm{FDT_{\mrm{DMD}}}^{-1} f_i$) where the optimal values of $(r, \tau_{\infty}/{\tau_d}) = (20, 0.833)$ are used in (\ref{eqn:FDT}) (see the text). (d) Inverse problem: the forcing needed for a specified time-mean response $\langle \overline{\theta} \rangle_{\mrm{target}}$ (solid line), is calculated as  $- \mbfi{L}_\mrm{FDT_{\mrm{POD}}} \langle \overline{\theta} \rangle_\mrm{target}$ and $- \mbfi{L}_\mrm{FDT_{\mrm{POD}}} \langle \overline{\theta} \rangle_\mrm{target}$. To examine the accuracy, long DNS with these forcing is conducted and dashed/dotted lines show the calculated $\langle \overline{\theta} \rangle$.}
\label{fig:FDT}
\end{figure}

From the unforced system's DNS, after the flow reaches quasi-equilibrium, $N=1.1\times 10^5$ samples of $\overline{T}(z,t)-\langle \overline{T} \rangle(z)$ have been collected every $\Delta t=0.12\tau_{adv}$, where $\tau_{adv} = \sqrt{L_z/(g\alpha \Delta T)}$ is the advective timescale. As stated in \citet{Khodkar2018}, in this system $\tau_{adv}=0.4\tau_d= 0.0012\tau_{diff}$, where $\tau_d$ is the decorrelation time of the leading POD mode (POD1). Using this data, $\mbfi{L}_\mrm{FDT_{POD}}$ is calculated from (\ref{eqn:FDT}) for various values of $r$ and $\tau_\infty$. Several tests involving predicting the time-mean response to an external forcing or forcing needed for a specific time-mean response are used to evaluate the accuracy of the calculated $\mbfi{L}_\mrm{FDT_{POD}}$. The ``true'' responses or forcings for these tests are obtained using DNS of forced systems. Figure~\ref{fig:FDT} depicts the results for four of these tests.




\begin{figure}
  \centerline{\includegraphics[width=0.95\textwidth]{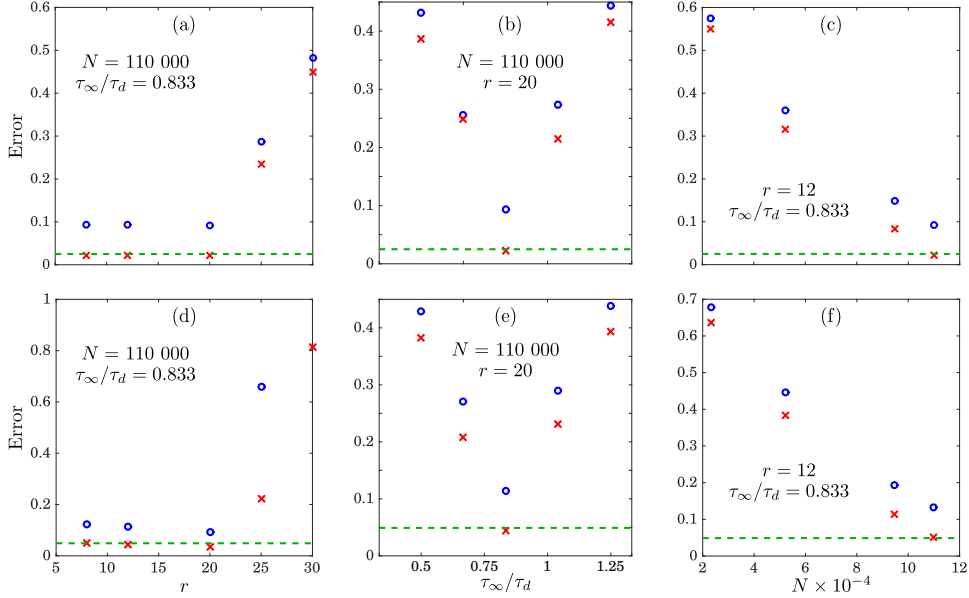}}
  \caption{The accuracy of $\mbfi{L}_\mrm{FDT}$ in predicting the response $\tm{\overline{\theta}}$ to forcings $f_1$ (top row) and $f_2$ (bottom row), as functions of the number of leading POD or DMD modes used in projection $r$ (left column), $\tau_{\infty}/\tau_d$ (middle column), and length of the dataset $N$ (right column). The values of other parameters which are maintained constant can be seen inside each panel. The blue circles and red crosses correspond to FDT$_\mrm{POD}$ and FDT$_\mrm{DMD}$, respectively. Since for the short datasets, the number of well-captured POD or DMD modes does not exceed $12$, we have used $r = 12$ in (c) and (f) for the sake of a fair comparison. The horizontal dashed lines show the errors of $\mbfi{L}_\mrm{GRF}$ from \citet{Khodkar2018}. Errors are calculated as $\norm{\tm{\overline{\theta}}_{\mbfi{L}} - \tm{\overline{\theta}}_{\mrm{DNS}}}/\norm{\tm{\overline{\theta}}_{\mrm{DNS}}}$.}
\label{fig:FDT_errors}
\end{figure}

As shown in figures~\ref{fig:FDT}(a--c), $\mbfi{L}_\mrm{FDT_{POD}}$ predicts the pattern of the time-mean responses well, but generally over- or under-estimates the amplitudes. Figure~\ref{fig:FDT}(d) demonstrates the accuracy of $\mbfi{L}_\mrm{FDT_{POD}}$ for the inverse problem (i.e., flow control): predicting the forcing needed to produce a specified change in the time-mean flow (i.e., a target response). As before, the FDT-predicted forcing can produce the pattern of the target reasonably well, but the amplitude is incorrect. The results presented in this figure are calculated using the optimal $(r,\tau_\infty)$, obtained from exploring the accuracy of the predicted responses/forcings in each case over a range of these two parameters (figure~\ref{fig:FDT_errors}). We find that for all tests, $(r=20,\tau_\infty=0.83 \tau_d)$ is optimal and leads to the closest agreement with the truth (i.e., DNS). These results suggest that for the best accuracy at $N=1.1\times 10^5$, independent of the forcing, the spatial dimension of the original samples $\bs{x}_t$ ($m=129$) should be reduced to $r=20$, and that $\tau_\infty$, which the accuracy is notably sensitive to, should be chosen slightly less than the decorrelation time of POD1 ($\tau_d$). The latter is consistent with the findings of \citet{gritsun2007climate} and \citet{Hassanzadeh2016b} in climate models. Figures~\ref{fig:FDT_errors}(c) and (f) show how the accuracy improves by increasing $N$. 

The results shown in figures~\ref{fig:FDT} and \ref{fig:FDT_errors} (and more tests, not shown) are promising, particularly given that the flow is complex and fully turbulent. However, the performance of $\mbfi{L}_\mrm{FDT_{POD}}$ is still not fullly satisfactory as the predicted amplitudes are inaccurate and the FDT$_\mrm{POD}$ is substantially outperformed by the accurate but computationally demanding (and not model-free) GRF method. {An analysis by \citet{Khodkar2018} based on the evaluation of the $\epsilon$-pseudospectrum showed that the RBC system under consideration here is moderately non-normal (see their Fig. 10b),}  which suggests that the performance of $\mbfi{L}_\mrm{FDT_{POD}}$ might be suffering from the same problem identified in \citet{Hassanzadeh2016b}: using the leading $r$ POD modes, which are orthogonal, can significantly degrade the performance of FDT$_\mrm{POD}$ if the system is non-normal, even if the $r \, (< m)$ modes explain a large percentage of the variance. This problem, and a potential remedy based on using DMD rather than POD modes for basis functions, is best seen by considering simple $2 \times 2$ systems of SDEs. This is done below, followed by applying FDT$_\mrm{DMD}$ to the same DNS dataset in \S\ref{sec:FDTDMD}. 



\subsection{Normal, Non-normal, and Nonlinear SDEs: FDT$_\mrm{POD}$ and FDT$_\mrm{DMD}$}



\begin{figure}
  \centerline{\includegraphics[width=0.95\textwidth]{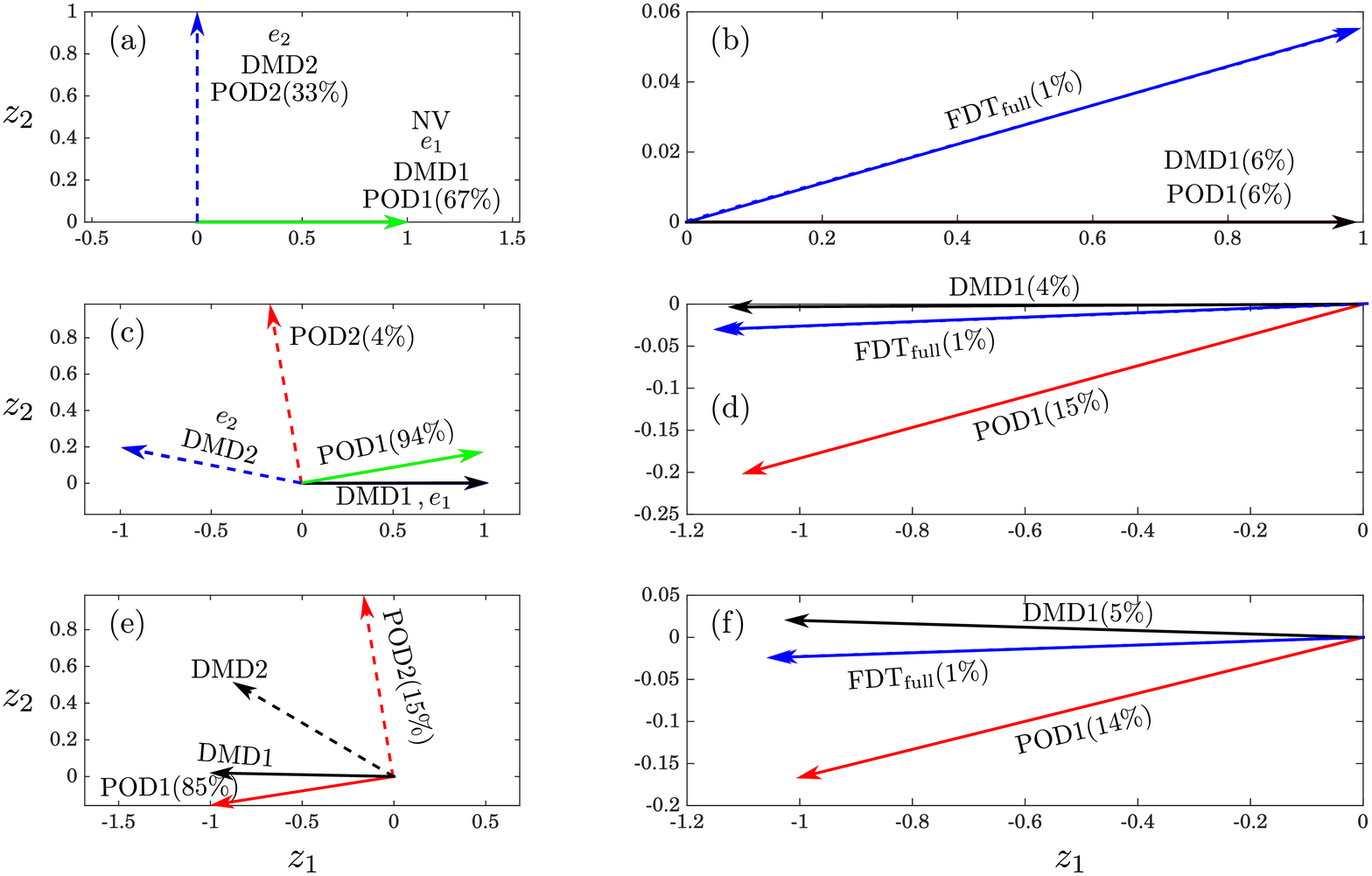}}
  \caption{Left column: eigenvectors $\bs{e}$ (blue lines), DMD modes (black lines), and POD modes (red lines) for the systems with $\mbfi{A}_1$ (top), $\mbfi{A}_2$ (middle), and $\mbfi{A}_3$ (bottom). $\bs{e}_1$ and DMD1 are the slower-decaying modes. Percentages show the variance explained by each POD mode. Right column: time-mean response to the forcing $(0.9\mrm{POD1} + 0.1\mrm{POD2})/\norm{0.9\mrm{POD1} + 0.1\mrm{POD2}}$, calculated using FDT (\ref{eqn:FDT}) with no projection (FDT$_\mrm{full}$, dashed blue lines), and projection onto the leading POD (FDT$_\mrm{POD1}$, red lines) or DMD (FDT$_\mrm{DMD1}$, black lines) mode. Solid blue lines show the analytically or numerically calculated true responses. Percentages show errors computed as $\norm{\bs{z}_{\mrm{FDT}} - \bs{z}_{\mrm{true}}}\times 100/\norm{\bs{z}_{\mrm{true}}}$. Note that the range of axes varies among panels.}
\label{fig:2D}
\end{figure}

We consider a two-dimensional SDE
\begin{equation}
	 \dot{\bs{z}} = \mbfi{A} \bs{z} + \bs{\xi} + \bs{f}    \, , \label{eqn:Theory9}
\end{equation}
where $\bs{z}^T = [z_1 \, z_2]$, $\bs{\xi}$ is Gaussian white noise, and $\bs{f}$ is a constant forcing. We use three test cases that are, respectively, normal, non-normal, and nonlinear with $\mbfi{A}$ being
\begin{equation}
\mbfi{A}_1 = \left[
	\begin{array}{cc}
		-1   & 0 \\
		 0   & -2 
	\end{array}
\right]  \, , \quad
\mbfi{A}_2 = \left[
	\begin{array}{cc}
		-1   &  5 \\
		 0   & -2 
	\end{array}
\right]  \, , \quad  
\mbfi{A}_3 = \left[
	\begin{array}{cc}
		-1   &  -4z_2 + 2 \\
		 0   &  -2 
	\end{array}
\right]  \, .  \label{eqn:Theory10}
\end{equation}
{We remark that the same matrices $\mbfi{A}_1$ and $\mbfi{A}_2$ were used in \citet{Hassanzadeh2016b} for a similar analysis which was focused only on POD modes.} Setting $\bs{f} = 0$, the SDE for each test case is integrated using the Euler-Maruyama method to generate datasets with 30s000 samples of $\bs{z}_t$. The POD and DMD modes are calculated from these datasets following \S\ref{section:Theory} and shown in figure~\ref{fig:2D}. As expected, for $\mbfi{A}_1$, the POD and DMD modes and eigenvectors are all identical (and each, orthogonal), while for $\mbfi{A}_2$, the DMD modes and eigenvectors are the same (and non-orthogonal) but different from the POD modes (which are orthogonal). For $\mbfi{A}_3$, the DMD and POD modes differ.




Time-mean responses to an external forcing $\bs{f}$ that is mostly in the direction of POD1 but has a small projection onto POD2 are predicted using $\mbfi{L}_\mrm{FDT}$ when no truncation is done (FDT$_\mrm{full}$), and when the data is truncated onto POD1 (FDT$_\mrm{POD1}$). For all test cases, FDT$_\mrm{full}$ has the error of $\sim 1\%$. While for the normal system FDT$_\mrm{POD1}$ is relatively accurate (error $\sim 6\%$), for the non-normal system the error is around $15 \%$ even though POD1 explains $94\%$ of the variance. To explain the source of this inaccuracy, following \citet{Hassanzadeh2016b}, we transfer (\ref{eqn:Theory9}) to the basis function space     
\begin{equation}
	 \dot{\bs{a}} = \left[ \mbfi{B}^{-1} \left( \mbfi{E} \, \mbfi{\Lambda} \, \mbfi{E}^{-1} \right) \mbfi{B} \right] \bs{a} + \mbfi{B}^{-1} \bs{f}    \, , \label{eqn:Theory11}
\end{equation}
where $\bs{a}^T = [a_1 \, a_2]$ are the projection coefficients, columns of $\mbfi{B}$ contain the basis functions (e.g., POD modes), $\mbfi{E}$ ($\mbfi{\Lambda}$) contain the eigenvectors (values) of $\mbfi{A}$, and $\bs{\xi}$ is ignored for convenience. For a normal system, the POD modes are the same as the eigenvectors and the matrix in the brackets reduces to the diagonal matrix $\mbfi{\Lambda}$, decoupling $a_1$ and $a_2$. Projections of $\bs{f}$ onto POD2 cannot be captured if $\mbfi{L}_\mrm{POD}$ is calculated only in the space of POD1; however, the accuracy of $a_1$ will not be affected, leading to the small error in figure~\ref{fig:2D}(b). For non-normal systems, the POD modes and eigenvectors can be significantly different, leading to a coupling between $a_1$ and $a_2$ that strengthens with non-normality \citep{Hassanzadeh2016b}. Hence, even small projections of $\bs{f}$ onto POD2 can substantially degrade the accuracy of $a_1$ (thus the FDT prediction) if $\mbfi{L}_\mrm{FDT}$ is calculated only in the space of POD1 (figure~\ref{fig:2D}(d)).                              

\begin{figure}
  \centerline{\includegraphics[width=0.95\textwidth]{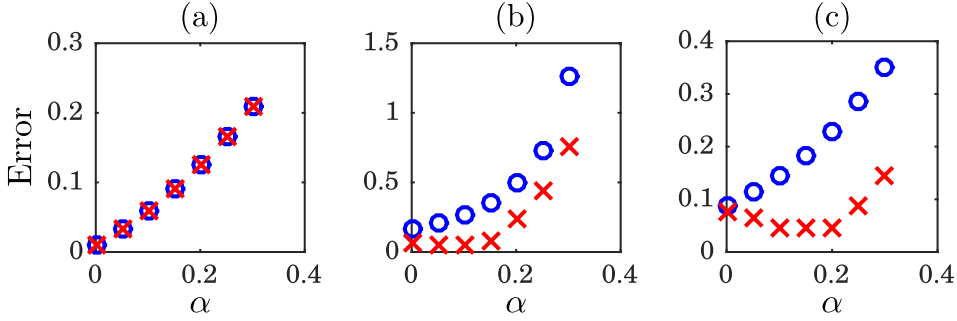}}
  \caption{For systems with (a) $\mbfi{A}_1$, (b) $\mbfi{A}_2$, and (c) $\mbfi{A}_3$, errors in time-mean response predictions to forcing $[(1- \alpha){\mrm{POD1}} + \alpha \mrm{POD2}]/\norm{(1- \alpha)\mrm{POD1} + \alpha \mrm{POD2}}$. Blue circles and red crosses indicate FDT$_\mrm{POD1}$ and FDT$_\mrm{DMD1}$, respectively. Error is computed as $\norm{\bs{z}_{\mrm{FDT}} - \bs{z}_{\mrm{true}}}/\norm{\bs{z}_{\mrm{true}}}$. Note that the range of $y$-axes varies among panels.}
\label{fig:2D_errors}
\end{figure}

The above analysis suggests that using basis functions that approximate the system's eigenvectors might improve the accuracy of $\mbfi{L}_\mrm{FDT}$. The discussion in \S\ref{section:Theory} and results in figure~\ref{fig:2D} point out to DMD modes as potential options. Indeed, using the slower-decaying DMD mode as the basis function (FDT$_\mrm{DMD1}$ hereafter) improves the accuracy compared to FDT$_\mrm{POD1}$ by a factor of four for the non-normal system. Similarly in the nonlinear system, the error of FDT$_\mrm{DMD1}$ is $5 \%$, three times lower than the $15 \%$ error of FDT$_\mrm{POD1}$. To further demonstrate the advantage of using the leading DMD rather than POD mode as basis function, figure~\ref{fig:2D_errors} shows that as the projection of the forcing onto POD2 increases in the non-normal and nonlinear systems, the accuracy of FDT-predicted responses rapidly degrades for FDT$_\mrm{POD1}$ while FDT$_\mrm{DMD1}$ shows a much better performance.    

\subsection{DNS of RBC at $Ra=10^6$: FDT$_\mrm{DMD}$}\label{sec:FDTDMD}
Using the unforced system's DNS, we have calculated{, following \S\ref{section:DMD} for $\tau=\Delta t$, all the $m=129$} DMD modes, and chosen the $r$ slowest-decaying ones as basis functions for FDT$_\mrm{DMD}$. If the $r^\mrm{th}$ DMD mode is complex, we ensure that its complex conjugate, also a DMD mode, is included in the basis function space as well. Figures~\ref{fig:FDT} and \ref{fig:FDT_errors} show that $\mbfi{L}_\mrm{FDT_{DMD}}$ accurately predicts the pattern and amplitude of the time-mean responses and remarkably outperforms $\mbfi{L}_\mrm{FDT_\mrm{POD}}$ in all cases. $\mbfi{L}_\mrm{FDT_{DMD}}$ has accuracy that is equal to (or in some cases better than) $\mbfi{L}_\mrm{GRF}$. Note that while DMD modes provide suitable basis functions for FDT, we have found that $\mbfi{L}_\mrm{DMD}$ (or $\mbfi{L}_\mrm{LIM})$ cannot accurately predict the time-mean responses/forcings for tests similar to those in figure~\ref{fig:FDT} (not shown).             

\color{black}

\section{Discussions and Conclusions \label{section:Conclusion}}

The DMD-enhanced FDT method is shown to accurately predict the time-mean response to an external forcing, or the forcing needed for a specified response in a canonical buoyancy-driven turbulent flow, RBC at $Ra=10^6$. Tests using the DNS of RBC and simple non-normal and nonlinear SDEs demonstrate the advantage in using a limited number of leading (slowest-decaying) DMD modes over the commonly used POD modes as basis functions for the FDT calculations in (\ref{eqn:FDT}). 

{This approach suggests a potential remedy to overcome the challenge identified by \citet{Hassanzadeh2016b} in applying FDT to high-dimensional, non-normal turbulent flows. Nonetheless, while the 1D linear ROM calculated using FDT$_\mrm{DMD}$ remarkably outperforms FDT$_\mrm{POD}$ and accurately predicts the pattern and amplitude of time-mean responses/forcings for the Rayleigh-B\'enard turbulence considered here, how the DMD-enhanced FDT performs for computing 2D and 3D ROMs and for more complex, higher-dimensional turbulent systems remains to be examined in future work. Large-scale atmospheric circulation, for which FDT$_\mrm{POD}$ has been extensively attempted with mixed outcomes, will be a test case of particular interest.}

{Finally, given the non-normality of the system, the sensitivity of the calculated responses to initial condition in initial-value integrations of Eq.~(\ref{eqn:Theory7}) and to small changes in the forcing (in both DNS and (\ref{eqn:Theory7})) should be investigated in subsequent studies.}

\color{black}

\section*{Acknowledgment \label{section:Acknowledgment}}

We thank Thanos Antoulas, Hessam Babaee, Mohammad Farazmand, Piyush Grover, Matthias Heinkenschloss and Saleh Nabi for insightful discussions, three anonymous reviewers for helpful comments, NSF grant AGS-1552385, NASA grant 80NSSC17K0266, and Rice University's Faculty Initiative Fund for support, and XSEDE Stampede2 (via allocation ATM170020) and Rice's DAVinCI cluster for providing computing resources.

\color{black}



\bibliographystyle{jfm}
\bibliography{Main}

\end{document}